%% file: cdf_dmix_v3.tex
\documentclass[aps,prl,twocolumn,superscriptaddress,showpacs]{revtex4}
\usepackage{graphicx}	% Include figure files
\usepackage{dcolumn}	% Align table columns on decimal point
\usepackage{bm}		% bold math

\begin{document}

\def\babar{\mbox{\sl B\hspace{-0.4em} {\small\sl A}\hspace{-0.37em} \sl B\hspace{-0.4em} {\small\sl A\hspace {-0.02em}R}}}

\title{
Evidence for $D^0 - \bar{D}^0$ Mixing Using the CDF~II Detector}

\input{July_2007_Authors_Visitors}
\date{\today}

\begin{abstract}

We measure the time dependence of the ratio of decay rates for the
rare decay $D^0 \rightarrow K^+\pi^-$ to the Cabibbo-favored decay
$D^0 \rightarrow K^-\pi^+$.   We use a signal of $12.7 \times 10^3$ $D^0
\rightarrow K^+\pi^-$ decays with proper decay times between 0.75 and
10 mean $D^0$ lifetimes. The data sample was recorded with the CDF~II
detector at the Fermilab Tevatron and corresponds to an integrated
luminosity of 1.5 fb$^{-1}$ for $p\bar{p}$ collisions at
$\sqrt{s}$ = 1.96 TeV.  We search for $D^0 - \bar{D}^0$ mixing and
measure the mixing parameters to be $R_D = (3.04 \pm
0.55) \times 10^{-3}$, $y' = (8.5 \pm 7.6) \times 10^{-3}$, and $x'^2 =
(-0.12 \pm 0.35) \times 10^{-3}$.
We report Bayesian probability contours in the $x'^2-y'$ plane and
find that the data are inconsistent with the no-mixing hypothesis with
a probability equivalent to 3.8 Gaussian standard
deviations.
 
\end{abstract}


\maketitle

Since the discovery of the charm quark in 1974 \cite{ref:BNL-J,
ref:SLAC-psi}, physicists have been searching for the oscillation of
neutral charm mesons between particle and anti-particle states.
Such behavior is referred to as ``mixing'',
as first explained in 1955~\cite{ref:K0mixing} for the $K^0$ meson
in terms of quantum-mechanical mixed states. Mixing was next observed
for $B^0_d$ mesons in 1987~\cite{ref:UA1_Bd_mixing, ref:ARGUS_Bd_mixing}.
The years 2006 and
2007 have seen landmark new results on mixing: observation of $B_s$
mixing from the CDF experiment \cite{ref:Bs_mixing} and evidence
for $D^0$ mixing from the Belle~\cite{ref:Belle_mixing} and
\babar~\cite{ref:BaBar_mixing} experiments.

The recent evidence for $D^0$ mixing comes from two different types of
measurements. 
The Belle Collaboration found direct evidence for a longer and shorter
lived $D^0$ meson, in analogy to the well-known case for $K^0$
mesons. They found significantly different decay time distributions
for $D^0$ decays to the CP-eigenstates $K^+K^-$ and $\pi^+\pi^-$
compared to that for the CP-mixed state $K^-\pi^+$.
(In this Letter, reference to a specific decay chain implicitly
includes the charge-conjugate decay.) No other experiment has
confirmed the evidence for lifetime differences among these
decays. The evidence for $D^0$ mixing found in the
\babar~experiment is a
difference in decay time distribution for $D^0 \rightarrow K^+\pi^-$
compared to that for the Cabibbo-favored (CF) decay $D^0 \rightarrow
K^-\pi^+$.  Such a difference depends on the combined effects of
differences in the masses and lifetimes of the $D^0$ meson weak
eigenstates.
This same measurement was made in the Belle experiment~\cite{ref:Belle_dcs},
but evidence for mixing was not seen.
In this Letter, we present a new measurement of the same $D^0$ mixing
process as used by
\babar~for their evidence.

In the standard model, the decay $D^0 \rightarrow K^+\pi^-$ proceeds
through a doubly Cabibbo-suppressed (DCS) ``tree'' diagram, and may
also result from a mixing process ($D^0 \leftrightarrow \bar{D^0}$),
if it exists, followed by a CF decay ($\bar{D^0} \rightarrow
K^+\pi^-$).  The DCS decay rate depends on Cabibbo-Kobayashi-Maskawa
quark-mixing matrix elements and on the magnitude of SU(3) flavor symmetry
violation~\cite{ref:Gronau-Rosner}.  Mixing may occur through two
distinct types of second-order weak processes. In the first, the $D^0$
evolves into a virtual (``long-range'') intermediate state such as
$\pi^+\pi^-$, which subsequently evolves to a $\bar{D}^0$.  The
magnitude of the amplitude for long-range mixing 
has been estimated using strong interaction models
~\cite{ref:FGLNP-2004}, but has not been determined using 
a QCD calculation from first principles.  The second type of
second-order weak process is short-range \cite{ref:Golowich_2005},
with either a ``box'' or ``penguin'' topology.  Short-range mixing is
negligible in the standard model. However, exotic weakly interacting
particles could enhance the short-range mixing and provide a signature
of new physics
\cite{ref:Burdman-Shipsey-2003,ref:GPP-2007,ref:GHPP-2007}.

The ratio $R$ of $D^0 \to K^+\pi^-$ to $D^0 \to K^-\pi^+$ decay rates can be
approximated \cite{ref:Blaylock-mix, ref:cleo} as a simple quadratic function
of $t/\tau$, where $t$ is the proper decay time and $\tau$ is the mean $D^0$
lifetime.  This form is valid assuming CP
conservation and small values for the parameters $x = \Delta M / \Gamma$
and $y = \Delta \Gamma / 2 \Gamma$, where $\Delta M$ is the mass
difference between the $D^0$ meson weak eigenstates, $\Delta \Gamma$
is the decay width difference, and $\Gamma$
is the average decay width of the eigenstates.
Under the assumptions stated above,
\begin{eqnarray}
R(t/\tau) = R_D + \sqrt{R_D} {y}^{\prime} \,(t/\tau) + { x^{\prime 2} + 
         y^{\prime 2} \over 4} \, (t/\tau)^2 ,
\label{eqn:R(t)}
\end{eqnarray}
where $R_D$ is the squared modulus of the ratio of DCS to CF amplitudes. 
The parameters $x^{\prime}$ and $y^{\prime}$ are linear combinations of $x$ and
$y$ according to the relations
$$
x^{\prime} = x ~{\rm cos}~\delta + y ~{\rm sin} ~\delta ~~~{\rm and}~~~ 
y^{\prime} = - x ~{\rm sin} ~\delta + y ~{\rm cos} ~\delta ,
$$
where $\delta$ is the strong interaction phase difference between the
DCS and CF amplitudes.  In the absence of mixing, $x' = y' = 0$ and $R(t/\tau) = R_D$.

Our measurement uses data collected by the CDF~II detector at the
Fermilab Tevatron collider, from February 2002 to January 2007,
corresponding to an integrated luminosity of $\approx
1.5$~fb$^{-1}$ for $p\bar{p}$ collisions at $\sqrt{s}=1.96$~TeV.
CDF~II \cite{ref:CDF} is a multi-purpose detector with a magnetic
spectrometer surrounded by a calorimeter and a muon detector.  The
detector components pertinent to this analysis are the silicon
microstrip vertex detector, the multi-wire drift
chamber (COT), and the 1.4~T magnet which together
measure the trajectories and momenta of charged particles.
The COT measures ionization energy loss for a charged particle, which is used
for particle identification (PID).

Events are selected in real time with a trigger system developed for a
broad class of heavy flavor decays.  The trigger requirements used
here are the same as those described for our previous
measurement of the time-integrated value of
$R$~\cite{ref:CDF-RB}, which used
a smaller data sample. The trigger selects events with a pair of oppositely charged
particles that are consistent with originating from a secondary
decay vertex separated from the beamline.

In the off-line analysis, we reconstruct the ``right-sign" (RS) CF
decay chain $D^{*+} \to \pi^+ D^0$, $D^0 \to K^- \pi^+$, and the
``wrong-sign" (WS) decay chain $D^{*+} \to \pi^+ D^0$, $D^0 \to K^+
\pi^-$.  The relative charges of the pions determine whether the decay
chain is RS (like charge) or WS (opposite charge).  The reconstruction
method is similar to that used for our previous time-independent
measurement.  The RS and WS $D^*$ decays have the same kinematics, and
differ only in decay time distributions.  To reduce systematic
uncertainties, we use the same selection criteria (cuts) for both the
RS and WS decay modes.  Analysis cuts were optimized before the WS
candidates were revealed, and were chosen to maximize the expected WS
signal significance.

The $D^0$ candidate reconstruction starts with a pair of tracks from
oppositely charged particles that satisfy the trigger requirements.
The tracks are considered with both $K^-\pi^+$ and $\pi^-K^+$
interpretations.  A third ``tagging'' track, required to have $p_T
\geq 0.3$~GeV/$c$, is used to form a $D^*$ candidate when considered
as a pion and combined with the $D^0$ candidate.

We apply two cuts to reduce the background to the WS signal from RS
decays where the $D^0$ decay tracks are misidentified because the kaon
and pion assignments are mistakenly interchanged.
As determined from the data, 96.4\% of $D^0$ decays with correct mass
assignment are reconstructed with $K\pi$ invariant mass $m_{K\pi}$
within 20 MeV/$c^2$ of the $D^0$ mass.
The $m_{K\pi}$ distribution for misidentified $D^0$ decays is much broader, and
has only 22\% of the events within the same mass range.
We remove WS candidates that have a RS mass within that range.
This cut excludes 96.4\% of RS decays and retains 78\% of the WS signal.
We also impose a cut based on PID which is used to distinguish pions from
kaons for all three tracks in the decay chain.  This cut, described
in Ref.~\cite{ref:CDF-RB}, further helps to reject misidentified decays.

We use a series of cuts based on the decay topology of signal events
in which a $D^{*}$ is produced at the primary vertex, the tagging
pion also originates from the primary vertex, and the $D^0$ travels a
measurable distance before decay.
The vertex-based cuts reduce combinatoric background from improper
combinations involving one or more tracks that do not originate from
the $D^*$ decay chain of interest.
We require the transverse decay length significance
$L_{xy}/\sigma_{xy}$ to be greater than 4, where $L_{xy}= \vec{r} \cdot \vec{p}_T / p_T$,  
$\vec{r}$ is the distance between the primary and $D^0$ decay vertices,
$\vec{p}_T$ is the transverse component of the momentum
of the $D^0$ candidate with respect to the beamline, and
$\sigma_{xy}$ is the uncertainty on $L_{xy}$.
The tagging pion track must have $d_0<500~\mu$m,
where the transverse impact parameter $d_0$ is the distance of
closest approach between a track and the
primary vertex in the plane transverse to the beamline.
The tagging pion must also have a point of closest approach to the primary vertex
less than 1.5 cm along the beamline.

The ratio $t/\tau$ is determined for each $D^0$ candidate by $t/\tau =
m_{D^0} L_{xy} / (p_T \tau)$, where $m_{D^0} = 1.8648$~GeV/$c^2$
and $\tau = 410.1$~fs are the world
average values for the $D^0$ invariant mass and lifetime,
respectively~\cite{ref:PDG}.  To study $R(t/\tau)$, we divide the data
into 20 bins of $t/\tau$ ranging from 0.75 to 10.0, choosing bins of
increasing size from 0.25 to 2.0 to reduce statistical uncertainty at
larger times.  The bin sizes are larger than the $t/\tau$
resolution of $\approx 0.16$.

\begin{figure}[tbp]
\begin{center}
\includegraphics[width=2.8in]{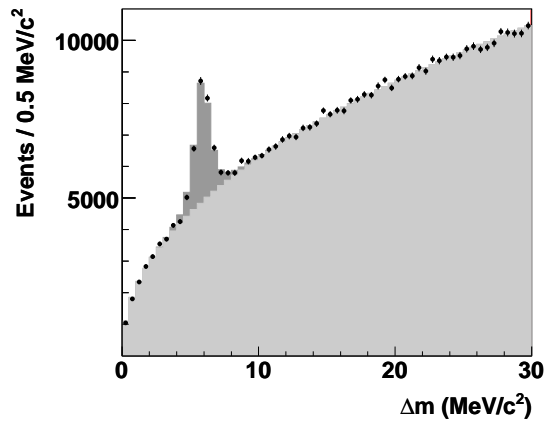}
\caption{Time-integrated distribution for ``wrong-sign'' $D^0 \rightarrow 
K^+ \pi^- $ signal yield as a function of $\Delta m$. Also shown is the result
of a least-squares fit using an empirical function for the
signal (dark shaded region) and a power law for the background (light
shaded region). }
\label{fig:WS_mass_diff}
\end{center}
\end{figure}

After RS and WS candidates are separately divided into $t/\tau$ bins, they are
further divided into bins of mass difference
$\Delta m \equiv m_{K\pi\pi} - m_{K\pi} -m_{\pi}$. For each $\Delta m$ bin,
we perform a binned maximum likelihood fit of the
corresponding $m_{K\pi}$ distribution to determine the $D^0$
signal yield.  The distribution of $D^0$ signal yield versus $\Delta m$ is fit
using a least-squares method to get the $D^*$ signal for each time bin.
The $D^*$ fit procedure is illustrated by the time-integrated WS $\Delta m$ distribution
shown in Fig. \ref{fig:WS_mass_diff}.

The signal shapes for the individual $m_{K\pi}$ and $\Delta m$ distributions
are fixed from the RS time-integrated fits.  For each $m_{K\pi}$
distribution, a parabola with floating parameters is used to fit the
background. The background shapes for all the $\Delta m$ WS (RS)
distributions are fixed to the shape determined for the
time-integrated WS (RS) distribution.  The amplitudes of the signal and background
shapes are determined independently for all $m_{K\pi}$ and $\Delta m$ fits.
The RS distributions have similar amounts of background as the WS distributions,
but the RS signal is about 250 times larger.

\begin{figure}[tbp]
\begin{center}
\includegraphics[width=2.8in]{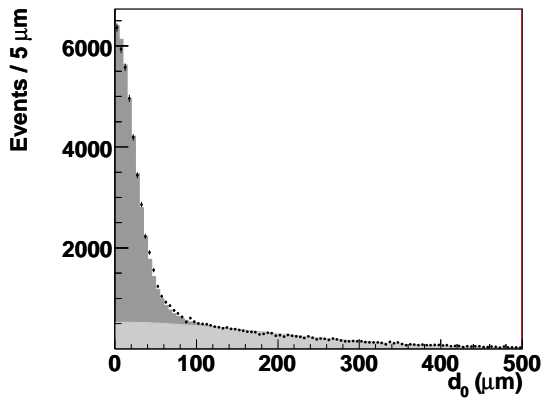}
\caption{
The distribution of transverse impact parameter $d_0$
for $5 < t/\tau < 6$ for ``right sign'' background-subtracted $D^*$ mesons.
The result of a binned maximum likelihood fit shows the
narrow peak due to promptly produced $D^*$ mesons (dark shaded)
and the broad distribution due to non-prompt $D^*$ mesons from $B$
decay (light shaded).}
\label{fig:ip_dist}
\end{center}
\end{figure}

The $D^*$ mesons that originate from beauty hadron ($B$) decays
must be treated as background to avoid the complication of measuring
the $D^0$ decay length from the $B$ decay point instead of the primary vertex.
The $D^*$ mesons produced promptly at the primary vertex
have a narrow $d_0$ distribution, with a shape
independent of $t/\tau$.
The background from non-prompt $D^*$ mesons from the decay chain
$B \rightarrow D^* \rightarrow D^0$ have a broad $d_0$ distribution, due to
the decay length of the $B$ hadrons.  The width of the broad distribution
increases with increasing $t/\tau$.
An example $d_0$ distribution is shown in Fig. \ref{fig:ip_dist}.
The shapes of the prompt and broad distributions are determined from RS data.
The WS shapes are the same as the RS shapes.
For each of the 20 $t/\tau$ bins, the prompt WS (RS) signal
is determined from the number of WS (RS) $D^*$ mesons and the
shapes of the $d_0$ distributions.  The ratio of
non-prompt to prompt signal is $\approx 0.02$ at $t/\tau = 2$
and increases with increasing $t/\tau$ due to the faster exponential fall-off with
$t/\tau$ for $D^0$ compared to $B$.  At $t/\tau = 7$, the ratio is $\approx 1$.

\begin{figure}[tbp]
\begin{center}
\includegraphics[width=2.8in]{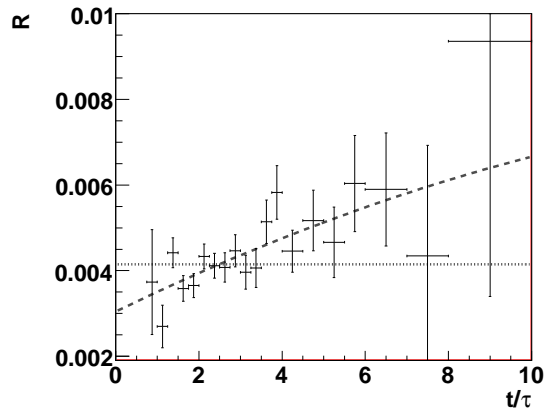}
\caption{Ratio of  prompt $D^*$ ``wrong-sign'' to ``right sign'' decays as 
a function of normalized proper decay time.  The dashed curve is from a 
least-squares parabolic fit, which determines the parameters $R_D$, $y'$, and 
$x'^2$.  The dotted line is the fit assuming no mixing.}
\label{fig:ratio_data}
\end{center}
\end{figure}

The time-integrated prompt $D^*$
signals are $(12.7 \pm 0.3) \times 10^3$ WS events and
$(3.044 \pm 0.002) \times 10^6$ RS events.
The ratios of prompt WS to RS signal for the 20 $t/\tau$ bins are shown in
Fig.~\ref{fig:ratio_data}. The uncertainties for each bin include
statistical and systematic contributions. The significant systematic uncertainties
are due to the background shapes for the $m_{K\pi}$, $\Delta m$, and $d_0$
distributions, which are described by parameters that are allowed to
vary in the fitting procedure.  We used simulation to confirm that our
choice of decay time bins does not systematically affect the result.
The detector acceptances for RS and WS decays are nearly identical, and their
difference contributes a negligible systematic uncertainty in the ratio $R$. 
The large uncertainty in the first $t/\tau$ bin
is due to low signal statistics caused by the trigger turn-on at small
$t/\tau$.  In the last two bins, the uncertainties are large because the
exponential fall-off of the WS signal with $t/\tau$ results in 
smaller numbers of signal events.

A least-squares parabolic fit of the data in Fig.~\ref{fig:ratio_data}
to Eq.~(\ref{eqn:R(t)}) determines the values and uncertainties for
the parameters $R_D$, $y'$, and $x'^2$, which are listed in
Table~\ref{tab:parameters}.  Since the value of $x'^2$ is unphysical
(less than zero), but consistent with zero, we also fit the data with
the constraint $x'^2$ = 0. The values of $R_D$ and $y'$ are consistent
with and without the constraint. The values and precision of
the parameters measured by CDF are comparable to those from the best
previous measurements, as shown in Table~\ref{tab:compare}.

\begin{table}[tbp]
 \caption[Mixing Parameters From Fits of the $R(t/\tau)$ Distribution]
 {Fit results for the $R(t/\tau)$ distribution. 
The uncertainties include statistical and systematic components.
The correlation coefficient between $y'$ and
$x'^2$ for the unconstrained fit is $-0.98$.
The no-mixing fit is consistent with our previous time-independent result
 \cite{ref:CDF-RB}.
}
\begin{center}
    \begin{tabular}{lcccc}
      \hline \hline
       Fit type
       & $R_D (10^{-3})$ & $y' \, (10^{-3})$ & $x'^2 \, (10^{-3})$
      & $\chi^2$ / d.o.f. \\
\hline
Unconstrained  &  3.04 $\pm$ 0.55 & 8.5 $\pm$ 7.6 & $-0.12$ $\pm$ 0.35   & 19.2 / 17 \\
Physically &  &  &  & \\
allowed  &  3.22 $\pm$ 0.23 & 6.0 $\pm$ 1.4 & 0  & 19.3 / 18 \\
No mixing    &  4.15 $\pm$ 0.10 & 0 & 0 & 36.8 / 19 \\
      \hline \hline
    \end{tabular}
 \end{center}
\label{tab:parameters}
\end{table}

\begin{table}[tbp]
 \caption[Comparison of the 
CDF Result With Recent Measurements]
 {Comparison of the CDF result with recent measurements.
All results use $D^0 \rightarrow K^+\pi^-$ decays and fits assuming 
no CP violation.
The uncertainties include statistical and systematic components.
The significance for no mixing is given in terms of the equivalent number
of Gaussian standard deviations.}
\begin{center}
    \begin{tabular}{lcccc}
      \hline \hline
       &  &  &  & Mixing  \\
      Experiment  & $R_D (10^{-3})$ & $y' \, (10^{-3})$ & $x'^2 \, (10^{-3})$
      & Signif. \\
\hline
CDF & 3.04 $\pm$ 0.55 & 8.5 $\pm$ 7.6 & $-0.12$ $\pm$ 0.35 & 3.8 \\

\babar~\cite{ref:BaBar_mixing}  & 3.03 $\pm$ 0.19 &
 9.7 $\pm$ 5.4 & $-0.22$ $\pm$ 0.37 & 3.9 \\

Belle~\cite{ref:Belle_dcs}  & 3.64 $\pm$ 0.17 & 0.6 $^{+4.0}_{-3.9}$ &
 ~0.18 $^{+0.21}_{-0.23}$ & 2.0 \\
      \hline \hline
    \end{tabular}
 \end{center}
\label{tab:compare}
\end{table}

\begin{figure}[tbp]
\begin{center}
\includegraphics[width=2.8in]{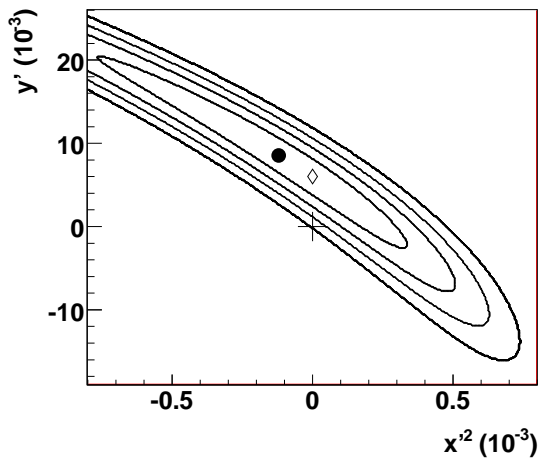}
\caption{Bayesian probability  contours in the $x'^2 - y'$ parameter 
space corresponding to one through four equivalent Gaussian standard
deviations. The closed circle shows the unconstrained fit values for the mixing parameters.
The open diamond shows the values from the physically allowed fit ($x'^2 \ge 0$).
The cross shows the no-mixing point.}
\label{fig:param_contour}
\end{center}
\end{figure}

To determine the consistency of our data with the no-mixing
hypothesis, we compute Bayesian contours containing the region with
the highest posterior probability.  The probability density is
calculated as the product of a likelihood ${\cal L}$ and a prior,
divided by a normalization factor.  The likelihood is ${\cal L} =
\exp(-\chi^2/2)$, where $\chi^2$ is computed from the data in
Fig.~\ref{fig:ratio_data} for a particular set of fit parameters.  The
normalization factor is the integral of ${\cal L}$ over the parameter
space.  A flat prior is used for all three parameters, and
$R_D$ is treated as a Bayesian nuisance parameter.  The contours
are insensitive to modest changes in the prior.  The contours
in the $x'^2$-$y'$ plane are shown in
Fig. \ref{fig:param_contour}. The no-mixing point lies on the contour
which excludes a region containing a probability of $1.5 \times
10^{-4}$, equivalent to 3.8 Gaussian standard deviations.  We also
computed contours with the constraint $x'^2 \ge 0$ and find a
probability for no-mixing consistent with the value obtained without
the constraint.

We tried alternate procedures to determine the probability for no
mixing. We fit the data in Fig.~\ref{fig:ratio_data} with the
constraint $y' = x'^2 = 0$, with results as given in
Table~\ref{tab:parameters}.  The change in log likelihood ($-2\Delta \ln
{\cal L}$) between the unconstrained and no-mixing fits has an
approximately chi-square distribution for two degrees of
freedom. From Table~\ref{tab:parameters},
$-2\Delta \ln {\cal L} = 17.6$, which corresponds to a probability of
$1.6 \times 10^{-4}$.  We also made a frequentist check using
ensembles of simulated $R(t/\tau)$ measurements without mixing.  The
probability for a simulation to have a value of $-2\Delta \ln {\cal L}
\ge 17.6$ is $1.3 \times 10^{-4}$.
The probabilities from both of these
checks are consistent with that obtained using Bayesian contours.

In conclusion, our data shows evidence for $D^0 - \bar{D}^0$ mixing in
the $K^+\pi^-$ channel, providing the first confirmation of the
evidence from the \babar~experiment. The mixing could be due to standard model 
long-range intermediate states or due to new physics. Improved reliability
of standard model calculations and future measurements of mixing signatures
with improved precision are needed to explain this phenomenon.

We thank the Fermilab staff and the technical staffs of the
participating institutions for their vital contributions. This work
was supported by the U.S. Department of Energy and National Science
Foundation; the Italian Istituto Nazionale di Fisica Nucleare; the
Ministry of Education, Culture, Sports, Science and Technology of
Japan; the Natural Sciences and Engineering Research Council of
Canada; the National Science Council of the Republic of China; the
Swiss National Science Foundation; the A.P. Sloan Foundation; the
Bundesministerium f\"ur Bildung und Forschung, Germany; the Korean
Science and Engineering Foundation and the Korean Research Foundation;
the Science and Technology Facilities Council and the Royal
Society, UK; the Institut National de Physique Nucleaire et Physique des
Particules/CNRS; the Russian Foundation for Basic Research; the Comisi\'on
Interministerial de Ciencia y Tecnolog\'{\i}a, Spain; the European Community's
Human Potential Programme; the Slovak R\&D Agency; and the Academy of Finland.

\end{document}

%% file: July_2007_Authors_Visitors.tex
\affiliation{Institute of Physics, Academia Sinica, Taipei, Taiwan 11529, Republic of China} 
\affiliation{Argonne National Laboratory, Argonne, Illinois 60439} 
\affiliation{Institut de Fisica d'Altes Energies, Universitat Autonoma de Barcelona, E-08193, Bellaterra (Barcelona), Spain} 
\affiliation{Baylor University, Waco, Texas  76798} 
\affiliation{Istituto Nazionale di Fisica Nucleare, University of Bologna, I-40127 Bologna, Italy} 
\affiliation{Brandeis University, Waltham, Massachusetts 02254} 
\affiliation{University of California, Davis, Davis, California  95616} 
\affiliation{University of California, Los Angeles, Los Angeles, California  90024} 
\affiliation{University of California, San Diego, La Jolla, California  92093} 
\affiliation{University of California, Santa Barbara, Santa Barbara, California 93106} 
\affiliation{Instituto de Fisica de Cantabria, CSIC-University of Cantabria, 39005 Santander, Spain} 
\affiliation{Carnegie Mellon University, Pittsburgh, PA  15213} 
\affiliation{Enrico Fermi Institute, University of Chicago, Chicago, Illinois 60637} 
\affiliation{Comenius University, 842 48 Bratislava, Slovakia; Institute of Experimental Physics, 040 01 Kosice, Slovakia} 
\affiliation{Joint Institute for Nuclear Research, RU-141980 Dubna, Russia} 
\affiliation{Duke University, Durham, North Carolina  27708} 
\affiliation{Fermi National Accelerator Laboratory, Batavia, Illinois 60510} 
\affiliation{University of Florida, Gainesville, Florida  32611} 
\affiliation{Laboratori Nazionali di Frascati, Istituto Nazionale di Fisica Nucleare, I-00044 Frascati, Italy} 
\affiliation{University of Geneva, CH-1211 Geneva 4, Switzerland} 
\affiliation{Glasgow University, Glasgow G12 8QQ, United Kingdom} 
\affiliation{Harvard University, Cambridge, Massachusetts 02138} 
\affiliation{Division of High Energy Physics, Department of Physics, University of Helsinki and Helsinki Institute of Physics, FIN-00014, Helsinki, Finland} 
\affiliation{University of Illinois, Urbana, Illinois 61801} 
\affiliation{The Johns Hopkins University, Baltimore, Maryland 21218} 
\affiliation{Institut f\"{u}r Experimentelle Kernphysik, Universit\"{a}t Karlsruhe, 76128 Karlsruhe, Germany} 
\affiliation{Center for High Energy Physics: Kyungpook National University, Daegu 702-701, Korea; Seoul National University, Seoul 151-742, Korea; Sungkyunkwan University, Suwon 440-746, Korea; Korea Institute of Science and Technology Information, Daejeon, 305-806, Korea; Chonnam National University, Gwangju, 500-757, Korea} 
\affiliation{Ernest Orlando Lawrence Berkeley National Laboratory, Berkeley, California 94720} 
\affiliation{University of Liverpool, Liverpool L69 7ZE, United Kingdom} 
\affiliation{University College London, London WC1E 6BT, United Kingdom} 
\affiliation{Centro de Investigaciones Energeticas Medioambientales y Tecnologicas, E-28040 Madrid, Spain} 
\affiliation{Massachusetts Institute of Technology, Cambridge, Massachusetts  02139} 
\affiliation{Institute of Particle Physics: McGill University, Montr\'{e}al, Canada H3A~2T8; and University of Toronto, Toronto, Canada M5S~1A7} 
\affiliation{University of Michigan, Ann Arbor, Michigan 48109} 
\affiliation{Michigan State University, East Lansing, Michigan  48824} 
\affiliation{University of New Mexico, Albuquerque, New Mexico 87131} 
\affiliation{Northwestern University, Evanston, Illinois  60208} 
\affiliation{The Ohio State University, Columbus, Ohio  43210} 
\affiliation{Okayama University, Okayama 700-8530, Japan} 
\affiliation{Osaka City University, Osaka 588, Japan} 
\affiliation{University of Oxford, Oxford OX1 3RH, United Kingdom} 
\affiliation{University of Padova, Istituto Nazionale di Fisica Nucleare, Sezione di Padova-Trento, I-35131 Padova, Italy} 
\affiliation{LPNHE, Universite Pierre et Marie Curie/IN2P3-CNRS, UMR7585, Paris, F-75252 France} 
\affiliation{University of Pennsylvania, Philadelphia, Pennsylvania 19104} 
\affiliation{Istituto Nazionale di Fisica Nucleare Pisa, Universities of Pisa, Siena and Scuola Normale Superiore, I-56127 Pisa, Italy} 
\affiliation{University of Pittsburgh, Pittsburgh, Pennsylvania 15260} 
\affiliation{Purdue University, West Lafayette, Indiana 47907} 
\affiliation{University of Rochester, Rochester, New York 14627} 
\affiliation{The Rockefeller University, New York, New York 10021} 
\affiliation{Istituto Nazionale di Fisica Nucleare, Sezione di Roma 1, University of Rome ``La Sapienza," I-00185 Roma, Italy} 
\affiliation{Rutgers University, Piscataway, New Jersey 08855} 
\affiliation{Texas A\&M University, College Station, Texas 77843} 
\affiliation{Istituto Nazionale di Fisica Nucleare, University of Trieste/\ Udine, Italy} 
\affiliation{University of Tsukuba, Tsukuba, Ibaraki 305, Japan} 
\affiliation{Tufts University, Medford, Massachusetts 02155} 
\affiliation{Waseda University, Tokyo 169, Japan} 
\affiliation{Wayne State University, Detroit, Michigan  48201} 
\affiliation{University of Wisconsin, Madison, Wisconsin 53706} 
\affiliation{Yale University, New Haven, Connecticut 06520} 
\author{T.~Aaltonen}
\affiliation{Division of High Energy Physics, Department of Physics, University of Helsinki and Helsinki Institute of Physics, FIN-00014, Helsinki, Finland}
\author{J.~Adelman}
\affiliation{Enrico Fermi Institute, University of Chicago, Chicago, Illinois 60637}
\author{T.~Akimoto}
\affiliation{University of Tsukuba, Tsukuba, Ibaraki 305, Japan}
\author{M.G.~Albrow}
\affiliation{Fermi National Accelerator Laboratory, Batavia, Illinois 60510}
\author{B.~\'{A}lvarez~Gonz\'{a}lez}
\affiliation{Instituto de Fisica de Cantabria, CSIC-University of Cantabria, 39005 Santander, Spain}
\author{S.~Amerio}
\affiliation{University of Padova, Istituto Nazionale di Fisica Nucleare, Sezione di Padova-Trento, I-35131 Padova, Italy}
\author{D.~Amidei}
\affiliation{University of Michigan, Ann Arbor, Michigan 48109}
\author{A.~Anastassov}
\affiliation{Rutgers University, Piscataway, New Jersey 08855}
\author{A.~Annovi}
\affiliation{Laboratori Nazionali di Frascati, Istituto Nazionale di Fisica Nucleare, I-00044 Frascati, Italy}
\author{J.~Antos}
\affiliation{Comenius University, 842 48 Bratislava, Slovakia; Institute of Experimental Physics, 040 01 Kosice, Slovakia}
% ...
\collaboration{CDF Collaboration\footnote{With visitors from $^a$University of Athens, 15784 Athens, Greece, 
$^b$Chinese Academy of Sciences, Beijing 100864, China, 
$^c$University of Bristol, Bristol BS8 1TL, United Kingdom, 
$^d$University Libre de Bruxelles, B-1050 Brussels, Belgium, 
$^e$University of California Irvine, Irvine, CA  92697, 
$^f$University of California Santa Cruz, Santa Cruz, CA  95064, 
$^g$Cornell University, Ithaca, NY  14853, 
$^h$University of Cyprus, Nicosia CY-1678, Cyprus, 
$^i$University College Dublin, Dublin 4, Ireland, 
$^j$University of Edinburgh, Edinburgh EH9 3JZ, United Kingdom, 
$^k$University of Heidelberg, D-69120 Heidelberg, Germany, 
$^l$Universidad Iberoamericana, Mexico D.F., Mexico, 
$^m$University of Manchester, Manchester M13 9PL, England, 
$^n$Nagasaki Institute of Applied Science, Nagasaki, Japan, 
$^o$University de Oviedo, E-33007 Oviedo, Spain, 
$^p$Queen Mary, University of London, London, E1 4NS, England, 
$^q$Texas Tech University, Lubbock, TX  79409, 
$^r$IFIC(CSIC-Universitat de Valencia), 46071 Valencia, Spain.}
}
\noaffiliation